\documentstyle[12pt]{article} 
\begin{document}
\begin{center}
{\bf REMARK ON CONSERVATION LAWS ASSOCIATED WITH NON-NOETHER SYMMETRIES}\\
\vspace*{5mm}
{\bf George Chavchanidze}

{\scriptsize Department of Theoretical Physics\\ 
A. Razmadze Institute Mathematics\\
1 Aleksidze Street, Ge 380093\\ 
Tbilisi, Georgia\\ e-mail:gch@rmi.acnet.ge}
\end{center}

\begin{abstract}
In the present paper geometric aspects of relationship
between non-Noether symmetries and conservation laws in Hamiltonian
systems is discussed. It is shown that integrals of motion associated with
continuous non-Noether symmetry are in involution whenever
generator of the symmetry satisfies certain Yang-Baxter equation.
\end{abstract}

In Hamiltonian systems time evolution of observables is governed by
Hamilton's equation
\begin{eqnarray} \label{eq:ham}
\frac{d}{dt}f = \{h , f\}
\end{eqnarray}
here $\{ , \}$ is Poisson bracket defined by some bivector field
$W$ satisfying condition
\begin{eqnarray} \label{eq:ww}
[W , W] = 0.
\end{eqnarray}
Namely
\begin{eqnarray}
\{f , g\} = W(df \wedge dg) = L_{W(f)}g = - L_{W(g)}f
\end{eqnarray}
where $W(f)$ and $W(g)$ are Hamiltonian
vector fields associated with $f$ and $g$ functions and
$L$ is Lie derivative.
Skew symmetry of bivector field $W$ provides skew symmetry of
Poisson bracket, while (\ref{eq:ww}) condition ensures Jacobi identity.
We also assume that dynamical system is regular - bivector field has maximal
rank (i. e. $W^{n} \neq 0$).
\\
Each vector field $E$ defines continuous one-parameter group
of transformations $g_{a} = e^{aL_{E}}$ that
acts on observables
\begin{eqnarray}
g_{a}(f) = e^{aL_{E}}(f) =
f + aL_{E}f + \frac{1}{2}a^{2}L_{E}^{2}f + ...
\end{eqnarray}
Such a group of transformation is a symmetry of (\ref{eq:ham}) Hamilton's equation
whenever its generator $E$ commutes with evolution operator
$W(h) = \{h , \}$ i. e.
\begin{eqnarray} \label{eq:sym}
[E , W(h)] = 0.
\end{eqnarray}
Symmetry is said to be Noether (or Cartan) if its generator is Hamiltonian
vector field (such a vector fields preserve $W$) and non-Noether
(non-Cartan) whenever generator is non-Hamiltonian (in other words
$[E , W] \neq 0$).
Non-Noether symmetries give rise to a number of integrals of motion
described in the following theorem \cite{lutzky}
(see also \cite{aut} \cite{autham} \cite{gascon}
\cite{hojman} \cite{mlutzky})
\\
{\bf Theorem:} If the vector field $E$ generates non-Noether
symmetry, then the functions
\begin{eqnarray} \label{eq:yl}
Y^{(l)} =
\frac{\hat{W}^{l}\wedge W^{n - l}}{W^{n}}
~~~~~l = 1,2, ... n
\end{eqnarray}
where $\hat{W} = [E , W]$ and $n$ is half dimension of the
phase space, are constant along solutions.
\\
{\bf Proof:}
By definition
\begin{eqnarray}
\hat{W}^{l}\wedge W^{n - l} = Y^{(l)}W^{n}
\end{eqnarray}
now let us take Lie derivative of this expression along the vector field $W(h)$.
\begin{eqnarray}
[W(h) , \hat{W}^{l}\wedge W^{n - l}] =
\frac{d}{dt}(Y^{(l)})W^{n}
+ Y^{(l)}[W(h) , W^{n}]
\end{eqnarray}
or
\begin{eqnarray}
l[W(h) , \hat{W}]\wedge \hat{W}^{l - 1}\wedge W^{n - l}
+ (n - l)[W(h) , W]\wedge \hat{W}^{l}\wedge W^{n - l - 1} = \nonumber \\
\frac{d}{dt}(Y^{(l)})W^{n} + nY^{(l)}[W(h) , W]\wedge W^{n - 1}
\end{eqnarray}
but according to Liouville theorem Hamiltonian vector field preserves $W$ i. e.
\begin{eqnarray}
[W(h) , W] = 0
\end{eqnarray}
and as a result
\begin{eqnarray}
[W(h) , \hat{W}] = [W(h)[E , W]] = [W[W(h) , E]] + [E[W , W(h)]] = 0
\end{eqnarray}
since $[W(h) , E] = 0$.
So $\frac{d}{dt}Y^{(l)} = 0$.
\\
{\bf Remark.} Instead of conserved quantities
$Y^{(1)} ... Y^{(n)}$
solutions $c_{1} ... c_{n}$ of the secular equation
\begin{eqnarray} \label{eq:sec}
(\hat{W} - cW)^{n} = 0
\end{eqnarray}
could be associated with generator of symmetry.
By expanding expression (\ref{eq:sec}) it is easy to verify that $Y^{(l)}$
integrals of motion can be expressed by means of $c_{1} ... c_{n}$
functions as follows
\begin{eqnarray} \label{eq:ylc}
Y^{(l)} = \frac{(n - l)! l!}{n!} \sum_{i_{p} \neq i_{s}} c_{i_{1}}c_{i_{2}} ... c_{i_{l}}
\end{eqnarray}
So $n$ integrals of motion are associated with each generator of non-Noether
symmetry. According to Liouville-Arnold theorem Hamiltonian system is
completely integrable if it possesses $n$ functionally independent integrals of
motion in involution (two functions $f$ and $g$ are said to be
in involution if their Poisson bracket vanishes $\{f , g\} = 0$).
Generally speaking conservation laws associated with symmetry might appear to be neither
independent nor involutive. However they are in involution if generator of the symmetry
satisfies Yang-Baxter type equation. Namely we have the following theorem.
\\
{\bf Theorem.}
If the vector field $E$ satisfies condition
\begin{eqnarray} \label{eq:yb}
[[E[E , W]]W] = 0
\end{eqnarray}
then the functions (\ref{eq:yl}) are in involution
\begin{eqnarray}
\{Y^{(k)} , Y^{(l)}\} = 0
\end{eqnarray} \\
{\bf Proof:} First of all let us note that identity (\ref{eq:ww}) satisfied by Poisson
bivector field $W$ is responsible for Liouville theorem
\begin{eqnarray}
[W , W] = 0 ~~~~~\leftrightarrow ~~~~~
L_{W(f)}W = [W(f) , W] = 0
\end{eqnarray}
By taking Lie derivative of the (\ref{eq:ww}) we obtain another useful identity
\begin{eqnarray}
L_{E}[W , W] = [E[W , W]] =
[[E , W] W] + [W[E , W]] =
2[\hat{W} , W] = 0
\end{eqnarray}
This identity gives rise to the following relation
\begin{eqnarray} \label{eq:com}
[\hat{W} , W] = 0 ~~~~~\leftrightarrow ~~~~~
[\hat{W}(f) , W] = - [\hat{W} , W(f)]
\end{eqnarray}
an finally condition (\ref{eq:yb}) ensures third identity
\begin{eqnarray}
[\hat{W} , \hat{W}] = 0
\end{eqnarray}
yielding Liouville theorem for $\hat{W}$
\begin{eqnarray}
[\hat{W} , \hat{W}] = 0 ~~~~~\leftrightarrow ~~~~~
[\hat{W}(f) , \hat{W}] = 0
\end{eqnarray}
Indeed
\begin{eqnarray}
[\hat{W} , \hat{W}] = [[E , W]\hat{W}] =
[[\hat{W} , E]W] = - [[E , \hat{W}]W] =
- [[E[E , W]]W] = 0
\end{eqnarray}
\\
Now let us consider two different solution $c_{i} \neq c_{j}$
of (\ref{eq:sec}) equation. By taking derivative of
\begin{eqnarray}
(\hat{W} - c_{i}W)^{n} = 0
\end{eqnarray}
equation along $W(c_{j})$ and
$\hat{W}(c_{j})$ vector fields and using Liouville theorem for
$W$ and $\hat{W}$ bivectors we obtain the following relations
\begin{eqnarray} \label{eq:cij}
(\hat{W} -
c_{i}W)^{n - 1}(L_{W(c_{j})}\hat{W}
- \{c_{j} , c_{i}\}W) =
0
\end{eqnarray}
and
\begin{eqnarray} \label{eq:cji}
(\hat{W} -
c_{i}W)^{n - 1}(c_{i}L_{\hat{W}(c_{j})}W
+ \{c_{j} , c_{i}\}_{\bullet }W) = 0
\end{eqnarray}
where
\begin{eqnarray}
\{c_{i} , c_{j}\}_{\bullet } =
\hat{W}(dc_{i} \wedge dc_{j})
\end{eqnarray}
is Poisson bracket calculated by means of $\hat{W}$ bivector field.
Now multiplying (\ref{eq:cij}) by $c_{i}$ subtracting (\ref{eq:cji}) and using
(\ref{eq:com}) identity gives rise to
\begin{eqnarray}
(\{c_{i} , c_{j}\}_{\bullet } -
c_{j}\{c_{i} , c_{j}\})(\hat{W} -
c_{i}W)^{n - 1}W = 0
\end{eqnarray}
So either
\begin{eqnarray} \label{eq:pbci}
\{c_{i} , c_{j}\}_{\bullet } -
c_{j}\{c_{i} , c_{j}\} = 0
\end{eqnarray}
or volume field
$(\hat{W} - c_{i}W)^{n - 1}W$
vanishes. In last case we can iterate (\ref{eq:cij}-\ref{eq:pbci}) procedure yielding after $n$
iterations $W^{n} = 0$ that according to our
assumption (dynamical system is regular) is not true.
As a result we have (\ref{eq:pbci}) and by simple exchange of indices $i \leftrightarrow j$
we get
\begin{eqnarray} \label{eq:pbcj}
\{c_{i} , c_{j}\}_{\bullet } -
c_{i}\{c_{i} , c_{j}\} = 0
\end{eqnarray}
Finally by comparing (\ref{eq:pbci}) and (\ref{eq:pbcj}) we obtain that
$c_{i}$ functions are in involution with respect to both
Poisson structures (since $c_{i} \neq c_{j}$)
\begin{eqnarray}
\{c_{i} , c_{j}\}_{\bullet } =
\{c_{i} , c_{j}\} = 0
\end{eqnarray}
and according to (\ref{eq:ylc}) the same is true for $Y^{(l)}$
integrals of motion.
\\
{\bf Note.}
Each generator of non-Noether symmetry satisfying equation (\ref{eq:yb}) endows dynamical system with bi- Hamiltonian
structure \cite{brouzet} \cite{autham} \cite{crampin} \cite{magri}
\cite{guha}
- couple ($W , \hat{W}$)
of compatible ($[W , \hat{W}] = 0$)
Poisson ($[W , W] = [\hat{W} , \hat{W}] = 0$)
bivector fields and give rise to bicomplex structure - \cite{crampin}\cite{adimakis}
\cite{dimakis} \cite{guha}
couple ($d = [W , ], \tilde{d} = [\hat{W} , ]$) of
compatible ($\tilde{d}d + d\tilde{d} = 0$) differential
($d^{2} = \tilde{d}^{2} = 0$) operators on graded algebra of
vector fields.
\\
{\bf Example:}
To illustrate correspondence between non-Noether symmetries and conservation
laws let us consider simple example
(other examples can be found in \cite{autham}) of particle experiencing action of dissipating force.
If friction depends linearly on velocity then time evolution in such a system
is governed by Hamilton's equation (\ref{eq:ham}) with
\begin{eqnarray}
W = \sum_{i = 1}^{n}
p_{i}\partial _{p_{i}} \wedge \partial _{q_{i}}
\end{eqnarray}
and
\begin{eqnarray}
h = \sum _{i = 1}^{n} p_{i} + q_{i}
\end{eqnarray}
And evolution operator is
\begin{eqnarray}
X = \sum _{i = 1}^{n}
p_{i}(\partial _{q_{i}} - \partial _{p_{i}})
\end{eqnarray}
This simple Hamiltonian system possesses nontrivial symmetry - it is invariant under the
following infinitesimal transformations
\begin{eqnarray}
q_{i} ~~~~~\rightarrow ~~~~~ q_{i} +
\epsilon (p_{i} + q_{i})^{2}
\end{eqnarray}
generated by
\begin{eqnarray}
E = \sum _{i = 1}^{n}
(p_{i} +q_{i})^{2}\partial _{q_{i}}
\end{eqnarray}
vector field. Clearly $E$ satisfies commutation relation (\ref{eq:sym}) and
Yang-Baxter equation (\ref{eq:yb}). Corresponding conserved quantities
\begin{eqnarray}
c_{i} = p_{i} + q_{i}
\end{eqnarray}
are in involution.
\\

\section{Acknowledgements}
{\it Author is grateful to Z. Giunashvili for
constructive discussions. This work was supported by INTAS (00-00561).}
\\

\end{document}